\documentclass[10pt,A4paper]{IEEEtran}

\usepackage{amsmath,epsfig,amssymb,verbatim,cite,amsopn}
\graphicspath{{/rasika/researchdocs/2005/jpapers/csup//xjournal/figs/}}
\newcommand{\pictwidth}{85truemm}
\newcommand{\E}{\mathrm{E}}

\newcommand{\nt}{n_t}
\newcommand{\nr}{n_r}

\newcommand{\del}{\beta}
\newcommand{\opvar}{\zeta}
\newcommand{\opvars}{\alpha}
\newcommand{\Pin}{P}

\newcommand{\vect}[1]{\boldsymbol{#1}}

\newcommand{\bX}{\vect{X}}

\newcommand{\bY}{\vect{Y}}

\newcommand{\bN}{\vect{N}}

\newcommand{\bH}{\vect{H}}

\title{Non-coherent Rayleigh fading MIMO channels: Capacity Supremum}

\author{Rasika R. Perera, Tony S. Pollock, Thushara D. Abhayapala\thanks{T. S. Pollock
and T. D. Abhayapala have appointments with National ICT Australia
(NICTA). Parts of this work has been approved at Asia Pacific
Conference on Communications (APCC), pp. 72-76, Oct. 2005, Perth,
Australia.}
\\Department of Information Engineering,\\Research
School of Information Sciences and Engineering,\\The Australian
National University, ACT 0200, AUSTRALIA.\\e-mail:
rasika.perera@anu.edu.au}

\begin{document}

\maketitle

\begin{abstract}
This paper investigates the limits of information transfer over a
fast Rayleigh fading MIMO channel, where neither the transmitter
nor the receiver has the knowledge of the channel state
information (CSI) except the fading statistics. We develop a
scalar channel model due to absence of the phase information in
non-coherent Rayleigh fading and derive a capacity supremum with
the number of receive antennas at any signal to noise ratio (SNR)
using Lagrange optimisation. Also, we conceptualise the discrete
nature of the optimal input distribution by posing the
optimisation on the channel mutual information for $N$ discrete
inputs. Furthermore, we derive an expression for the asymptotic
capacity when the input power is large, and compare with the
existing capacity results when the receiver is equipped with a
large number of antennas.
\end{abstract}


\section{Introduction}
Communication over rapidly time-varying channels where the
receiver is unable to estimate the channel state is a challenging
task. In particular, for a mobile receiver the estimation may
become difficult and the limits of information transfer in this
scenario is vital when the channel becomes non-coherent. In this
paper we study the capacity of the non-coherent Rayleigh fading
MIMO channel to identify the limits of information transfer with
no CSI at both ends.

One of the important problems in information theory is that of
computing the capacity of a given communication channel, and
finding the optimal input distribution that achieves capacity. The
capacity problem is addressed by maximising the mutual
information, a concave function subject to some constraints on the
channel input distribution. For discrete finite-size input
alphabets, this is a finite dimensional problem and all the input
constraints are such that they define a compact, convex set, where
the optimisation is straightforward. However, for continuous input
channels, neither the continuity of the objective function, nor
the compactness of the constraint set is now obvious, and the
optimisation for capacity is considered a difficult problem
\cite{{DTMRayleigh-Shami-2001},{Smith-1971}}.

A large amount of literature has appeared in the area of
multi-antenna wireless communications over the past decade.  Much
of the interest was motivated by the work in
\cite{{Telatar-1995},{Foschini-1999}} which shows a potential
increase in the channel capacity using multiple antennas for both
transmitting and receiving. The capacity of the MIMO channel, when
neither the transmitter nor the receiver has CSI, has
traditionally been considered an open and difficult problem. In
this paper, we provide an upper bound on the non-coherent Rayleigh
fading MIMO channel capacity for any receive antenna number at any
SNR and elaborate on the discrete nature of the optimal input.

The MIMO channel capacity derived in \cite{Telatar-1995}, is based
on the assumptions that i) the channel matrix elements are
independent and \textit{circularly symmetric}\footnote{The
distribution of a complex variable $\omega$ is said to be
circularly symmetric if for any deterministic
$-\pi\leq\,\theta\,\leq\,\pi$, the distribution of random variable
$e^{j\theta}{\omega}$ is identical to the distribution of
$\omega$.} complex Gaussian; ii) the noise at different receive
antennas are independent and white and; iii) the CSI is perfectly
known at the receiver but not to the transmitter. The main result
is linear capacity growth with the minimum number of transmit and
receive antennas. Also  \cite{Telatar-1995} shows that the optimum
input distribution which achieves this channel capacity is
circularly symmetric complex Gaussian which maximises the channel
output entropy when the CSI is known.

The capacity evaluation using Monte Carlo method when the receiver
CSI is perfectly known is shown in \cite{wang-chao-2003}. The
results extends Telatar's work \cite{Telatar-1995} and provides
accurate numerical results. Furthermore \cite{Mohammad-ali-2001}
shows the use of \textit{water filling} on the input Gaussian
vector when receiver and transmitter CSI is available at the both
transmitter and receiver. The increase in capacity is significant
for small number of antennas and at low SNR. At high SNR the
fading is less destructive and the water filling solution does not
provide a considerable gain over the capacity with CSI at the
receiver only.

The non-coherent channel capacity of the time selective Rayleigh
fading channel is studied in \cite{Yingbin-nocsi-corre-2004}.
Upper and lower bounds on the capacity at high SNR for single
antenna systems is shown, and derived a lower bound on the
capacity with multi-antennas. Abou-Faycal, Trott and Shamai
\cite{DTMRayleigh-Shami-2001} investigated the single input single
output (SISO) discrete time memoryless Rayleigh fading channel,
proving that the capacity achieving distribution is discrete with
a finite number of mass points. In addition, Taricco
\cite{Taricco-1997} showed the capacity supremum and confirmed
that the attainable input distribution is discrete in agreement
with Abou-Faycal's results. A general channel model is considered
in \cite{Meyn-2003} ignoring the structure of the channel.

Conversely in \cite{Smith-1971}, the channel is assumed to have
only additive noise, and in
\cite{DTMRayleigh-Shami-2001,Taricco-1997}, the channel is assumed
to have both additive and multiplicative noise. In
\cite{Meyn-2003} it is shown that under a peak power constrained
input, the capacity achieving input distribution is discrete, and
attempts to make a case for the argument that continuous capacity
achieving distributions are actually anomalies, and for most
channels the optimal input distribution is discrete with a finite
number of mass points.

Similar work is reported by Marzetta and Hochward
\cite{Marzetta-MIMO-1999} in non-coherent Rayleigh fading MIMO
channels using a block fading model over a coherence interval of
$T$ symbol periods. They characterised a certain structure of the
optimal input distribution and computed the channel capacity with
reduced complexity. The coherence capacity with channel coherence
time $T$ is used as the upper bound.

Also, it is shown that the non-coherent channel capacity
approaches the coherent capacity as $T$ becomes large where the
optimal input is approximately independent complex Gaussian. Zheng
and Tse \cite{Zheng-MIMO-2002} extended this work and specifically
computed the asymptotic capacity at both high and low SNR in terms
of $T$ and the number of transmit and receive antenna elements.
The suggested input is to transmit orthogonal vectors on a certain
number of transmit antennas with constant equal norms. They also
claim that having more transmit antennas than receive antennas
does not provide any capacity gain at high SNR, while having more
receive antennas yield a capacity gain.

Recently, Lapidoth and Moser \cite{Lapidoth-Moser-2002} showed the
capacity of non-coherent multi antenna systems grows only double
logarithmically in SNR and evaluated the \textit{fading
number}\footnote{The fading number is defined as the limit of the
difference between the channel capacity and
$\log(1+\log(1+\rho))$, where $\rho$ is the SNR.} with the optimum
input distributions in the high SNR region. This double
logarithmic bahavior is a good example to visualise the low
capacity available with no CSI compared to the coherent capacity
given for MIMO \cite{Telatar-1995,Foschini-1999} and SISO
\cite{Goldsmith-Varaiya-csi-1997} systems respectively.

In this paper we address the non-coherent uncorrelated Rayleigh
fading MIMO channel and show a capacity supremum optimising the
mutual information under average power constrained input for any
SNR.

The contributions of the paper as follows:
\begin{enumerate}
    \item We provide the mutual information of the non-coherent Rayleigh
fading MIMO channel in simple form using output differential
entropies.
    \item We optimise the mutual information using Lagrange
optimisation method and show a capacity supremum for a given
number of receive antennas at any SNR.
    \item We show the asymptotic
analysis of the capacity supremum with double logarithmic behavior
at high SNR, similar to the results shown in
\cite{Lapidoth-Moser-2002} and conjecture the discrete nature of
the optimal input.
    \item The proposed capacity in this paper can be
taken as an \textit{upper bound}\footnote{The term ``upper bound''
is used since for any input distribution either discrete or
continuous, the mutual information achieved through the channel is
lower than the capacity result derived in this paper.} to the
non-coherent uncorrelated MIMO channel capacity in Rayleigh
fading.
\end{enumerate}

The organisation of this paper is as follows. Section \ref{MM}
contains the channel model with notations used for the
non-coherent Rayleigh fading MIMO communication system. The
derivation of mutual information for the introduced channel model
is presented in Section \ref{MI}, along with the detailed work
based on Lagrange optimisation for channel capacity in
\ref{ncapacity}. Section \ref{ch4-NR} presents the numerical
results and analysis of our results. Finally conclusions are drawn
in Section \ref{con}.

\section{MIMO Channel Model}
\label{MM}

The input output relationship of a MIMO channel can be written as
\begin{equation}\label{mimo-model}
    \bY={\bH}{\bX}+{\bN},
\end{equation} where the output $\bY$ is $\nr\times1$, the channel gain
matrix $\bH$ is $\nr\times\nt$. The input $\bX$ is $\nt\times1$
and the noise $\bN$ which is assumed to be zero mean complex
Gaussian is $\nr\times1$. Each element of $\bH$, $h_{{i}{j}},
i=1,...,\nr, j=1,...,\nt$ is assumed to be zero mean circular
complex Gaussian random variables with a unit variance in each
dimension.

We use $X=|\bX|$ and $Y=|\bY|$ to denote the random scalar
variables where $|\cdot|$ is the Euclidean norm. $x$ and $y$
represent each realisation of $X$ and $Y$ (i.e. $x\,\epsilon\,X$
and $y\,\epsilon\,Y$). The input is power limited with an average
power constraint $\int{x^2}{p_{X}(x)}{d{x}}\leq P$. $\nt$ and
$\nr$ denote the number of transmit and receive antennas
respectively, and
$\gamma=-\int_{0}^{\infty}{e^{-y}}{\log{y}}{dy}\,\approx0.5772...,$\,
denotes the Euler's constant. We use $\Gamma(\cdot)$ and
$\Psi(\cdot)$ to indicate Gamma and Psi functions respectively.
$h(\bX)$ denotes the differential entropy of $\bX$, and
$I(\bX;\bY)$ designates the mutual information between $\bX$ and
$\bY$. The expected value over a set of random variables are
denoted by $E\{\cdot\}$, with $\text{det}\,(\cdot)$ for the
determinant, $(\cdot)^*$ for conjugate transpose of a matrix and
$I$ for an identity matrix. All the differential entropies and the
mutual information are defined to the base ``e'', and the results
are expressed in ``nats''. Neither the receiver nor the
transmitter has the knowledge of CSI except the fading statistics.
Channel coherence time is one where the channel changes
independently at every transmitted symbol.

\section{MIMO Mutual Information}
\label{MI}

\subsection{Capacity with Receiver CSI}
The Rayleigh fading MIMO channel capacity when the receiver has
the perfect CSI, given by,
\begin{equation}\label{telatar-capacity}
    C_{\text{rcsi}}=\E_{\bH}\left[\log{\text{det}}\left(I_{\nr}+\frac{\Pin}{\nt}{\bH}{\bH^*}\right)\right]
\end{equation}
was derived by Telatar in optimising the mutual information
$I(\bX;\bY)=h(\bY)-h(\bN)$ \cite{Telatar-1995} and later Foschini
\cite{Foschini-1999} who extended the work to show how the
capacity scales with increasing SNR for a large but practical
number of antenna elements at both the transmitter and receiver.
The linear growth of the capacity with $\nr$ is shown for a
special case where the channel matrix $\bH=I_{{\nr}{\nt}}$. The
capacity increase in \eqref{telatar-capacity} is more prominent
having multiple antennas at the receiver instead of the
transmitter. However, under fast fading conditions, the estimation
of fading coefficients which are assumed to be independent could
be difficult due to the short duration of fades. It is of interest
to study the capacity of such a channel and understand the
ultimate limits when no CSI is available. Furthermore, the
increasing demand for higher date rates along with mobility will
make the instantaneous channel measurement more difficult.
Therefore, it is important to find the optimal rate when the CSI
is not perfectly available at the receiver (non-coherent). In this
paper, we consider the mutual information of non-coherent
uncorrelated Rayleigh fading MIMO channel and investigate the
capacity.

\subsection{Mutual Information}
The conditional probability density function (pdf) of the output
given input of channel model \eqref{mimo-model} is given by
\begin{equation}\label{cond-pdf-real}
    f_{\bY|\bX}(y|x)=\frac{1}{(2{\pi}{(1+x^2)})^{\nr}}{\exp\left[{-{\frac{y^2}{2{(1+x^2)}}}}\right]},
\end{equation} where
$f_{\bY|\bX}(y|x)\triangleq\,f_{\bY|\bX}(y_1,y_2,...,y_{\nr}|x_1,x_2,...,x_{\nt})$,
$y\,\epsilon\,|\bY|$ and $x\,\epsilon\,|\bX|$. The magnitude sign
is removed in \eqref{cond-pdf-real} for simplicity and likewise in
the rest of this paper since the non-coherent Rayleigh fading
channel does not carry any phase information \cite{Taricco-1997}.
The pdf of the magnitude distribution of \eqref{cond-pdf-real} has
the form
\begin{equation}\label{cond-pdf-mag1}
    p_{Y|X}(y|x)=\frac{{y^{2{\nr}-1}}{\exp\left[{-{\frac{y^2}{2{(1+x^2)}}}}\right]}}
    {{2^{\nr-1}}{\Gamma(\nr)}{(1+x^2)^{\nr}}}
\end{equation} when Jacobian coordinate transformation is applied
on $2{\nr}$ dimensions. The output conditional entropy
$h(\bY|\bX)$ for \eqref{mimo-model} is given by
\begin{equation}\label{mimo-output-cond-entro1}
    h(\bY|\bX)=-\E_{x}\left\{{\int_{0}^{\infty}}{p_{Y|X}(y|x)}\,{\log\left[{p_{Y|X}(y|x)}\right]}{d{{y}}}\right\},
\end{equation} where the expectation is taken over
$x\,\epsilon\,X$. With \eqref{cond-pdf-mag1}, we get
\begin{align}
    h(\bY|\bX)&=\frac{1}{2}{\E_{x}\left\{{\log(1+x^2)}\right\}}+\log\left[\frac{\Gamma(\nr)}{\sqrt{2}}\right]\nonumber\\
    &-\left(\nr-\frac{1}{2}\right)\Psi(\nr)+\nr.
    \label{mimo-hyx-general-final}
\end{align}
Equation \eqref{mimo-hyx-general-final} can be used to calculate
the output conditional entropy of uncorrelated Rayleigh fading
MIMO channel when no CSI is available for a given input
distribution. With the output entropy
$h(\bY)=-\int_{0}^{\infty}{p_{Y}(y)}{\log[p_{Y}(y)]}{dy}$, we
obtain the mutual information \cite{Gallager-1968}
\begin{align}
I(\bX;\bY)&=h(\bY)-h(\bY|\bX)\nonumber\\
&=-\int_{0}^{\infty}{p_{Y}(y)}{\log \left[p_{Y}(y)\right]}{d{{y}}}-\frac{1}{2}{\E_{x}\left\{{\log(1+x^2)}\right\}}\nonumber\\
&-\log\left[\frac{\Gamma(\nr)}{\sqrt{2}}\right]+\left(\nr-\frac{1}{2}\right)\Psi(\nr)-\nr.
\label{ch4-mutual-information}
\end{align} From \eqref{ch4-mutual-information}, the mutual
information for a given input distribution can be computed.
However, finding the optimal input and hence the capacity for a
given input constraint is difficult. In next Section, we show how
to derive un upper bound on \eqref{ch4-mutual-information}
identifying some key properties of the optimal input.

\section{Non-Coherent MIMO Capacity}
\label{ncapacity}
\subsection{Output Constraints}
To obtain an expression for capacity of no-coherent Rayleigh
fading MIMO channel, equation \eqref{ch4-mutual-information} needs
to be maximised subject to an appropriate constraint on the input.
Usual constraints used are $\int_{0}^{\infty}{p_{X}(x)}{d{{x}}}=1$
and $\int_{0}^{\infty}{{x}^2}{p_{X}(x)}{d{{x}}}=\Pin$. However,
the maximisation of \eqref{ch4-mutual-information} subject to
these constraints does not provide a valid input pdf for SISO or
MIMO in this case. To overcome this difficulty, additional
constraints are used in \cite{Taricco-1997} for SISO non-coherent
Rayleigh fading channel. Likewise, to optimise
\eqref{ch4-mutual-information} we use the following constraints

\begin{subequations}
\begin{align}\label{py}
&\int_{0}^{\infty}{p_{Y}(y)}{d{{y}}}=1,
\end{align}
\begin{align}\label{yypy}
&\int_{0}^{\infty}{{y}^2}{p_{Y}(y)}{d{{y}}}={2}{\nr}(1+\Pin),
\end{align}
\begin{align}\label{constaints3}
&\int_{0}^{\infty}{p_{Y}(y)}{\log{y}}{d{{y}}}=\frac{1}{2}{({\del}+\Psi(\nr)+\log{2})},
\end{align}
\end{subequations} where $\beta={\E_{x}\left\{{\log(1+x^2)}\right\}}$. The second
constraint is the average mean squared power of $y\,\epsilon\,Y$,
which is considered as the induced power at the output by the
input, channel gain and noise. The constraint \eqref{constaints3},
is derived in Appendix \ref{constraint}. This additional
constraint is used to support the optimisation process in order to
arrive at a valid output pdf. Similar techniques are commonly
employed in convex optimisation work \cite{Boyd-2004}.

\subsection{Lagrange Optimisation}
Using the Lagrange variable $L$ and the multipliers $\lambda_{1},
\,\lambda_{2}$ and $\lambda_{3}$, we define
\begin{align}
L&=I(\bX;\bY)+\lambda_{1}{\left[\int_{0}^{\infty}{p_{Y}(y)}{d{{y}}}-1\right]}\nonumber\\
&+\lambda_{2}\left[\int_{0}^{\infty}
{{y}^2}{p_{Y}(y)}{d{{y}}}-{2}{\nr}(1+\Pin)\right]\nonumber\\
&+\lambda_{3}\left[\int_{0}^{\infty}{p_{Y}(y)}{\log{y}}{d{{y}}}-\frac{1}{2}{({\del}+\Psi(\nr)+\log{2})}\right].
\label{lagrangian}
\end{align}
Solving \eqref{lagrangian} for $p_{Y}(y)$, we obtain the optimum
output pdf
\begin{equation}\label{optimum-py-withlamdas}
p_{Y}(y)={\exp\left[{(\lambda_{1}-1)+\lambda_{2}{{y}^2}+\lambda_{3}{\log{y}}}\right]}
\end{equation}
for the mutual information in \eqref{ch4-mutual-information} in
terms of Lagrange variables.

We substitute optimum $p_{Y}(y)$ \eqref{optimum-py-withlamdas}
into three constraints \eqref{py}, \eqref{yypy}, and
\eqref{constaints3} and use the integral identities \cite[Page
360-365]{Table-of-integrals} to derive the following three
equations:
\begin{subequations}
\begin{equation}\label{app-const-1}
    \frac{(-\lambda_{2})^{-\left(\frac{1+\lambda_{3}}{2}\right)}}{2}
    {e^{\lambda_{1}-1}}{\Gamma\left(\frac{1+\lambda_{3}}{2}\right)}=1,
\end{equation}
\begin{align}
    \frac{(-\lambda_{2})^{-\left(\frac{3+\lambda_{3}}{2}\right)}}{2}
    {e^{\lambda_{1}-1}}{\Gamma\left(\frac{3+\lambda_{3}}{2}\right)}&=2{\nr}(1+\Pin),
    \label{app-const-2}
\end{align}
\begin{align}
    & \frac{e^{\lambda_{1}-1}{(-\lambda_{2})^{\frac{\lambda_{3}}{2}}}{\Gamma\left(\frac{1+\lambda_{3}}
    {2}\right)}}{4{\sqrt{-\lambda_{2}}}}
    \left\{\Psi\left(\frac{1+\lambda_{3}}{2}\right)-\log(-\lambda_{2})\right\}\nonumber\\
    &=\frac{1}{2}(\del+\Psi(\nr)+\log{2}),
    \label{app-const-3}
\end{align}
\end{subequations} where the second lagrange multiplier $\lambda_{2}$ is a negative
quantity. From \eqref{app-const-1} and \eqref{app-const-2}, and
using the relationship \cite[Page
255]{Hand-book-mathematical-functions}
\begin{equation}\label{gamma-relationship}
    \Gamma\left(\frac{3+\lambda_{3}}{2}\right)=\left({\frac{1+\lambda_{3}}{2}}\right)\Gamma\left(\frac{1+\lambda_{3}}{2}\right),\nonumber
\end{equation} we get
\begin{equation}\label{lamda-three}
    \lambda_{3}=-4{\lambda_{2}}{\nr}(1+\Pin)-1.
\end{equation} With \eqref{lamda-three} and expressing the quantity
$e^{\lambda_{1}-1}$ in terms of $\lambda_{2}$, we can solve
\eqref{app-const-3} for $\lambda_{2}$ with
\begin{align}
\Psi[-{2}{\lambda_{2}}{\nr}(1+\Pin)]-\log(-\lambda_{2})=\del+\Psi(\nr)+\log{2}.
\label{lamda-2-solution}
\end{align} Equation \eqref{lamda-2-solution} can be solved for $\lambda_{2}$ for certain $\nr$, $\Pin$
and $\del$ which is a function of $\Pin$ . We assume, there exists
a solution in the form
\begin{equation}\label{lamda2-final}
    \lambda_{2}=\frac{-\opvar}{2{\nr}(1+\Pin)},\,\,\,\,\,\,\,\,\opvar\,>\,0.
\end{equation} This assumption is made only to ease the rest of
the mathematics involved and has no effect on the capacity
results. From $\lambda_{2}$ we obtain
\begin{equation}\label{delta-and-optmised-variable}
    \del=\log\left\{\frac{\nr(1+\Pin)}{\opvar}\right\}+\Psi(\opvar)-\Psi(\nr),
\end{equation}
and following two equations for $\lambda_3$ and $\lambda_1$:
\begin{subequations}
\begin{equation}\label{}
    \lambda_{3}=2{\opvar}-1,
\end{equation}
\begin{equation}\label{}
    e^{\lambda_{1}-1}=\frac{{\opvar}^{\opvar}}{{{\nr}(1+\Pin)^{\opvar}}{\Gamma(\opvar)}}.
\end{equation}
\end{subequations} Substituting these Lagrange multipliers in
\eqref{optimum-py-withlamdas} we get the optimum output pdf
\begin{equation}\label{optimum-py}
    p_{Y}(y)=\frac{{{\opvar}^{\opvar}}{{y}^{{2}\opvar-1}}}{{{\nr}(1+\Pin)^{\opvar}}
    {\Gamma(\opvar)}}{\exp\left[{-\frac{\opvar{{y}^2}}{2{\nr}(1+\Pin)}}\right]}.
\end{equation}
\noindent\textit{We have following remarks:}
\begin{enumerate}
    \item To calculate capacity we need to find values for
    $\lambda_1$, $\lambda_2$ and $\lambda_3$.
    \item Equation \eqref{lamda-2-solution} could be used to
    calculate $\lambda_2$, for a given value of $\Pin$ and $\del$.
    However, it is difficult to find the value of
    $\beta={\E_{x}\left\{{\log(1+x^2)}\right\}}$ for a given P.
    \item Note that $0\,\leq\,\del\,\leq\,\log(1+\Pin)$, where the
    right inequality is derived by direct application of Jensen's
    inequality \cite{T-M-Cover-1991}.
    \item To better understand the characteristics of $\del$ we
    use \eqref{delta-and-optmised-variable} to plot both $\del$
    and $\del-\log(1+\Pin)$ vs $\opvar$ in Fig. \ref{beta-eta}. It is clear
    that $\del-\log(1+\Pin)$ is an increasing function of
    $\opvar$.
    \item In next Section we derive a supremum for the capacity
    using the properties of $\del$.
\end{enumerate}
\subsection{Capacity Supremum}
\label{supremum-section} 
\begin{figure} \centering
  \includegraphics[width=\pictwidth]{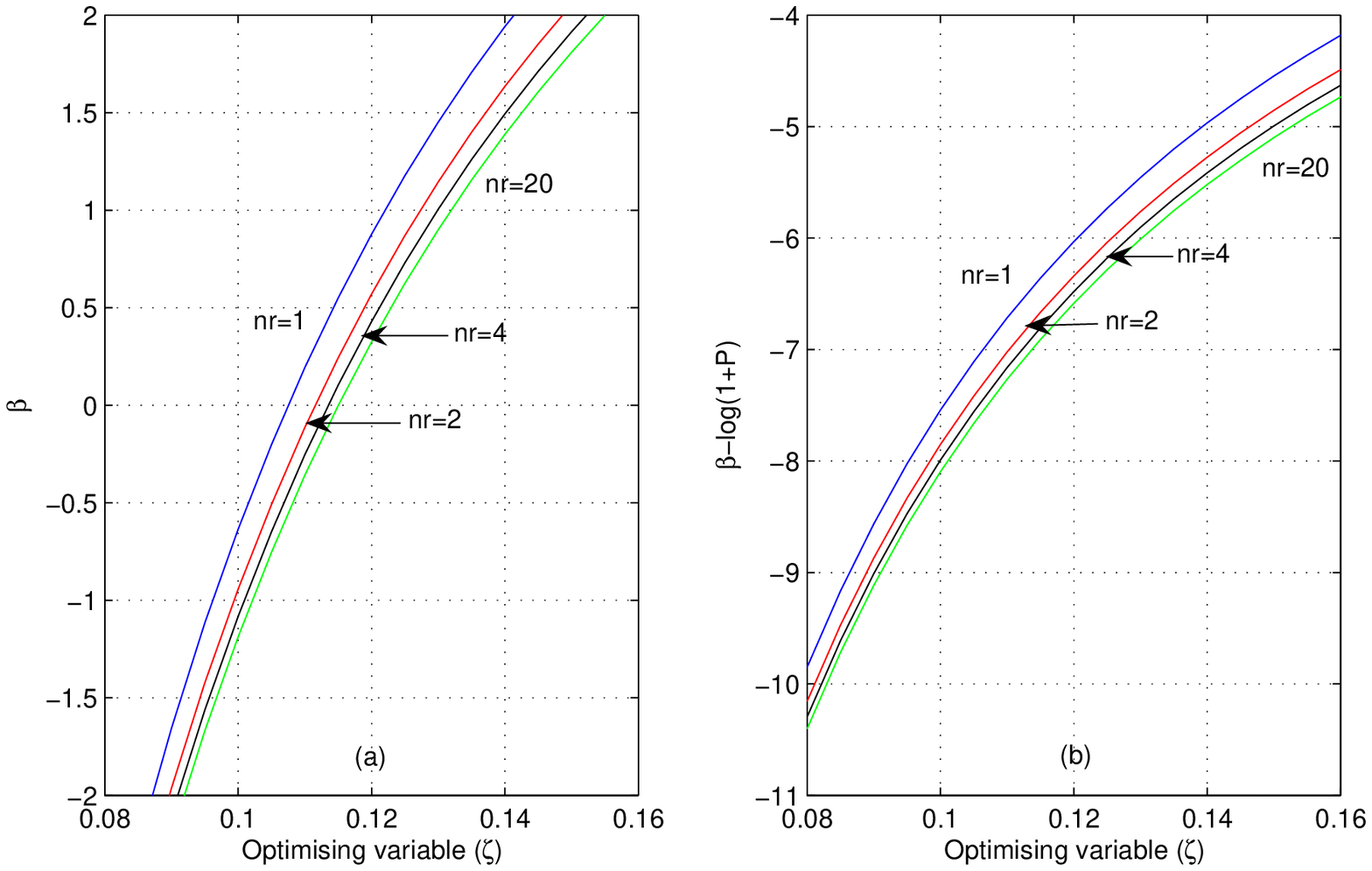}\\
  \caption{(a) The quantity $\del$ in
  \eqref{delta-and-optmised-variable} as a function of $\opvar$ showing
  $\opvar$ which produce both negative and positive values for
$\del=\E_{x}\{\log(1+x^2)\}$, SNR=30
  dB. (b) Shows $\del-\log(1+\Pin)$ is an increasing function of
  $\opvar$.}
\label{beta-eta}
\end{figure}

\begin{figure}
\centering
  \includegraphics[width=\pictwidth]{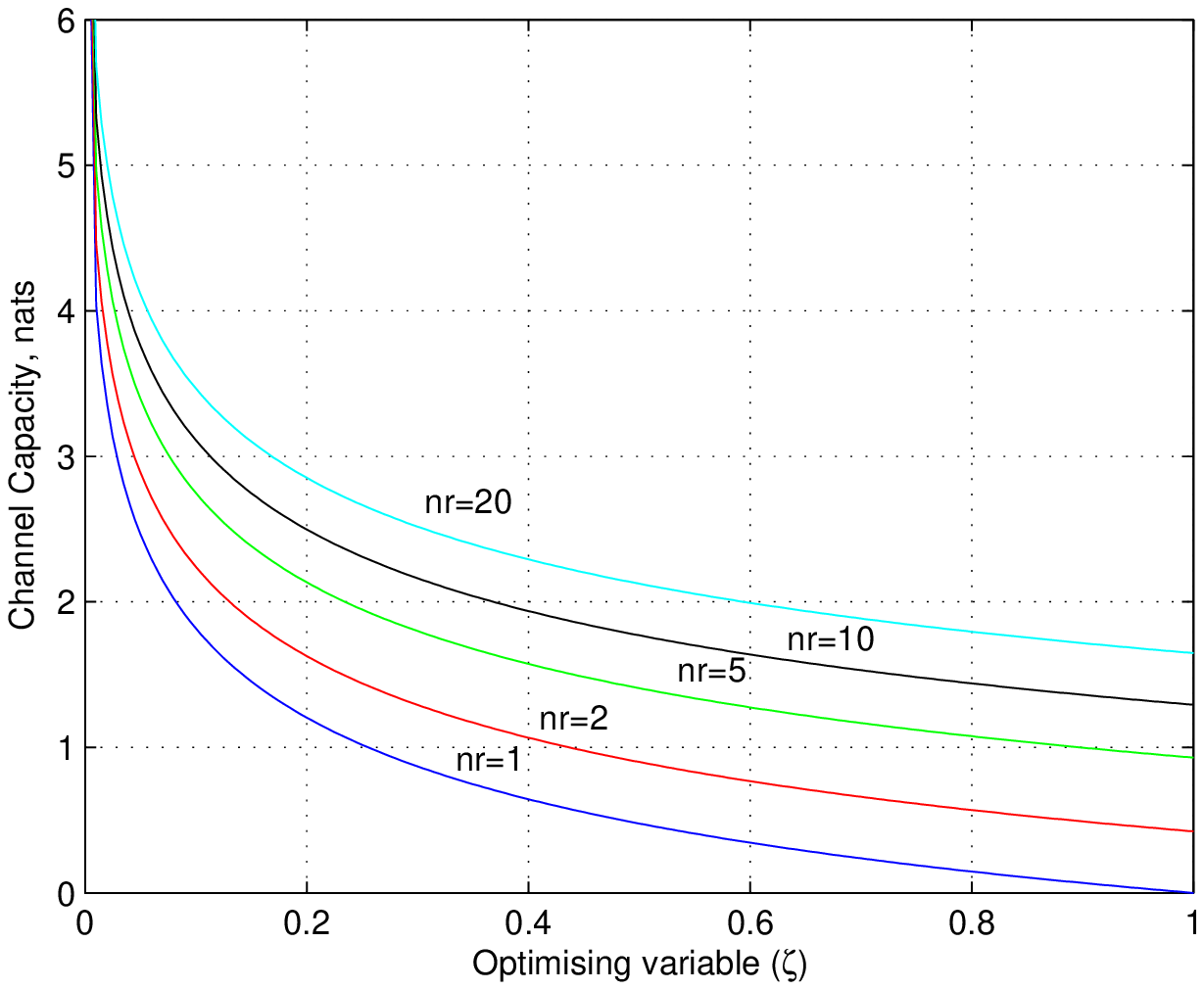}\\
  \caption{Channel capacity \eqref{sup-channel-capacity} as a function of
  $\opvar$, SNR=30dB.}\label{cap-eta}.
\end{figure}

Substituting the optimum $p_{Y}(y)$ from
\eqref{optimum-py} and $\del$ from
\eqref{delta-and-optmised-variable} into
\eqref{ch4-mutual-information}, we obtain the non-coherent channel
capacity
\begin{align}
C(\opvar)&=G(\opvar)-G(\nr),\label{sup-channel-capacity}
\end{align} where
\begin{equation}\label{ch4-G-function}
    G(\tau)=\log{\Gamma(\tau)}+\tau\left(1-\Psi(\tau)\right).
\end{equation}
From Fig. \ref{beta-eta} it is clear that low $\nr$ gives high
$\del$ for a given SNR. However, $\del\,\geq\,0$ since the
expectation over the function $\log(1+x^2)$ is always a positive
quantity. Therefore, the $\opvar$ which produce negative $\del$
values does not provide a valid capacity for a given $\nr$.
 The capacity $C(\opvar)$ is a monotonically
decreasing function of $\opvar$ since
\begin{equation}\label{}
    \frac{\partial{C(\opvar)}}{{\partial}{\opvar}}=-\opvar{\Psi_1(\opvar)},
\end{equation} where $\Psi_n(\cdot)$ is the $n^{th}$ derivative of
$\Psi(\cdot)$ \cite[page
253-255]{Hand-book-mathematical-functions}. Fig. \ref{cap-eta}
depicts the channel capacity for $\opvar\,\in\,(0,1)$. The
capacity can be computed for some $\del=\del'$, seeking the
optimal $\opvar$ which satisfies the input power constraint.
Furthermore, $\del$ in \eqref{delta-and-optmised-variable} is a
monotonically increasing function of $\opvar$ where
\begin{equation}\label{}
    \frac{\partial{\del(\opvar)}}{{\partial}{\opvar}}=-\frac{1}{\opvar}+\Psi_1(\opvar).
\end{equation} Therefore, the supremum of \eqref{sup-channel-capacity}
\begin{equation}\label{sup-capacity-supremum-one}
    C_{\text{sup}}=C(\opvar_s)=G(\opvar_s)-G(\nr)
\end{equation} is obtained with $\del=0$ where the corresponding
$\opvar_s$ is given by
\begin{equation}\label{input-power-1}
    \Psi(\opvar_s)-\log(\opvar_s)=\Psi(\nr)-\log\left[\nr(1+\Pin)\right].
\end{equation}

\begin{figure}
\centering
  \includegraphics[width=\pictwidth]{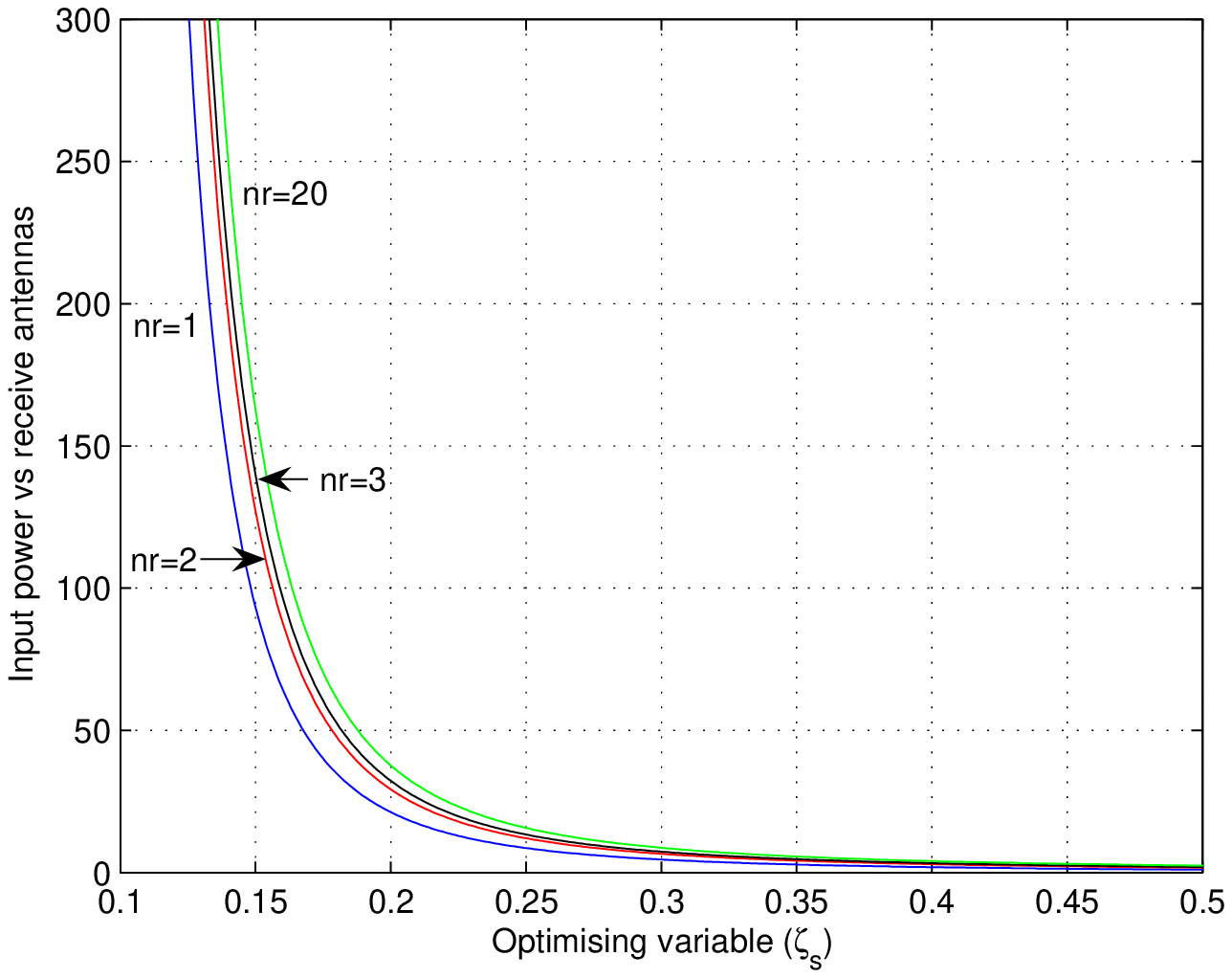}\\
  \caption{Input power \eqref{input-power-1} as a function of $\opvar_s$ showing the available solutions for $\opvar_s$
           for any $\Pin$, $\nr\,\in\,\{1,2,3,20\}$.}\label{power-eta}
\end{figure}

\begin{figure}
\centering
  \includegraphics[width=\pictwidth]{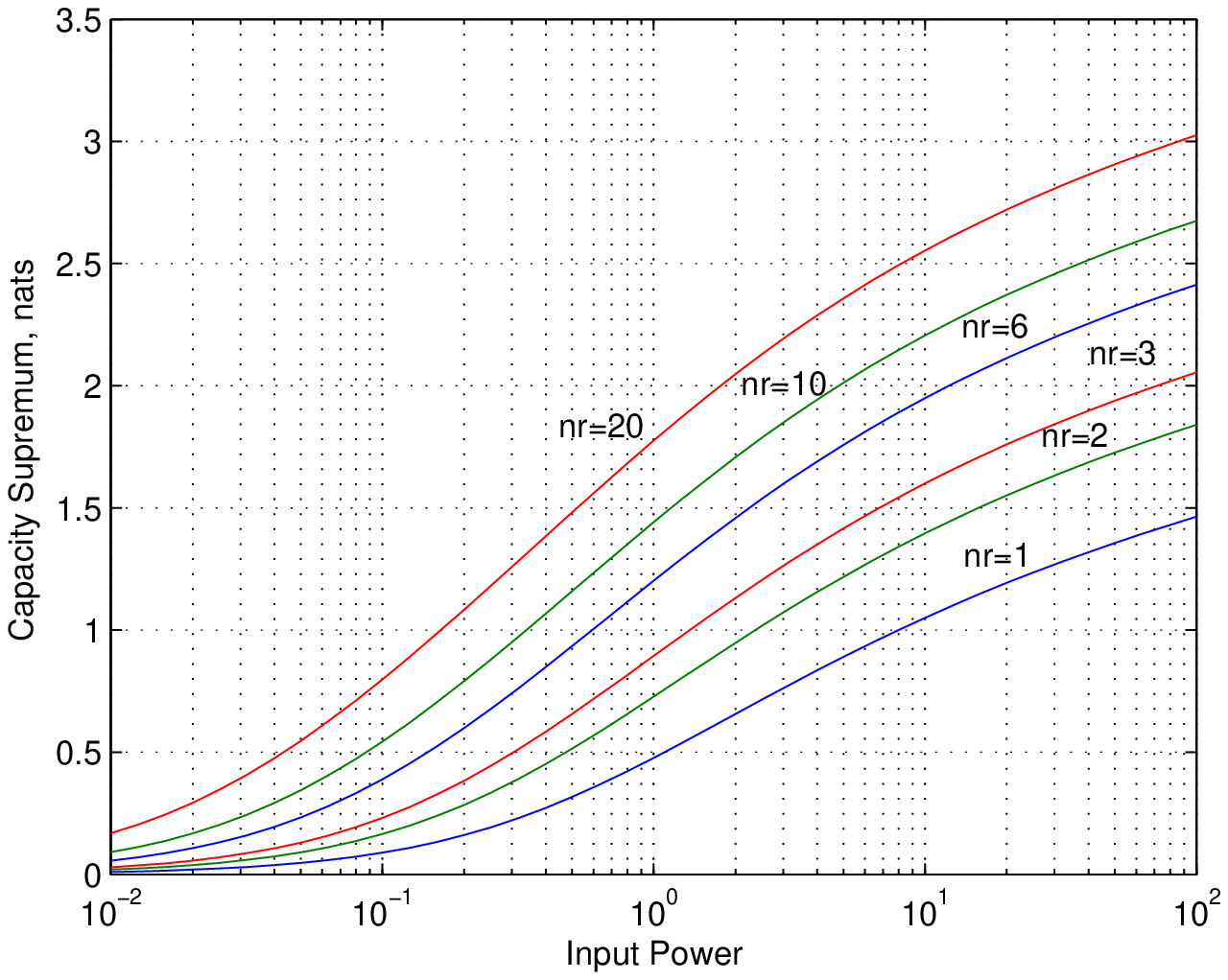}\\
  \caption{Capacity supremum \eqref{sup-capacity-supremum-one} of the non-coherent Rayleigh fading MIMO channel with the input power for
  different number of receive antennas $\nr$. }\label{cap-supremum}
\end{figure}

The input power $\Pin$ in \eqref{input-power-1} vs $\opvar_s$ as a
function of $\nr$ is given in Fig. \ref{power-eta}. It is clear
that there exist $\opvar_s$ which gives the solution to
\eqref{input-power-1} for any $\Pin$ for a given $\nr$.
 Fig. \ref{cap-supremum} shows the capacity supremum in
\eqref{sup-capacity-supremum-one} against the input power for
different $\nr$ equating $\opvar_s$ in \eqref{input-power-1}. This
non-coherent Rayleigh fading MIMO channel capacity supremum can be
used as an upper bound for all input distributions.

For larger $\nr$, Sengupta and Mitra \cite{Sengupta-2004} show the
capacity of non-coherent Rayleigh fading MIMO channel
\begin{equation}\label{ch4-capacity-singupta}
    C_{\text{MIMO}}=\frac{1}{2}{\log\left(\frac{\nr}{{2}{\pi}}\right)}+\log{\mu}+\frac{P}{\mu{\nt}\left(1+\frac{P}{\nt}\right)}{\,\,,}
\end{equation}
where
\begin{equation}\label{the-mu}
    \mu=\int_{0}^{\infty}\frac{d{y}}{1+y}{\exp\left[{-\frac{y}{\mu\left(1+\frac{P}{\nt}\right)}}\right]}\nonumber
\end{equation} and $\mu\approx\log(1+\frac{P}{\nt})$ for large $P$.
Furthermore, they established a continuous input distribution
which achieves \eqref{ch4-capacity-singupta} for large $\nr$. We
will discuss the analysis of the capacity supremum in
\eqref{sup-capacity-supremum-one} with respect to
\eqref{ch4-capacity-singupta} in Section \ref{ch4-NR}.

The capacity \eqref{sup-capacity-supremum-one} is independent of
the number of transmit antennas since the optimisation is carried
out using magnitude of the input vector. Therefore, the effect of
number of transmit antennas on capacity is not apparent. However,
\cite{Marzetta-MIMO-1999} proves that the capacity of $\nt\,>\,T$
is equal to the capacity for $\nt=T$ where $T$ is the channel
coherence time. In this paper, we consider $T=1$, and therefore
according to the \textit{theorem 1} in \cite{Marzetta-MIMO-1999},
the optimal $\nt=1$. Therefore, we conclude that capacity supremum
\eqref{sup-capacity-supremum-one} is true irrespective of $\nt$.
\subsection{Optimal Input Distribution}
The corresponding input distribution which provides this channel
capacity supremum \eqref{sup-capacity-supremum-one} for a certain
$\nr$ is given by
\begin{align}
    \int_{0}^{\infty}{p_{X}(x)}{\frac{{y^{2{\nr}-1}}{\exp\left[{-{\frac{y^2}{2{(1+x^2)}}}}\right]}}
    {{2^{\nr-1}}{\Gamma(\nr)}{(1+x^2)^{\nr}}}}{d{{x}}}&=
    \frac{{{\opvar_s}^{\opvar_s}}{{y}^{{2}\opvar_s-1}}}{{{\nr}(1+\Pin)^{\opvar_s}}{\Gamma(\opvar_s)}}\nonumber\\
    &\times\,{\exp\left[{-\frac{\opvar_s{{y}^2}}{2{\nr}(1+\Pin)}}\right]}.
\label{sup-input-distribution}
\end{align}

The integral in \eqref{sup-input-distribution} takes the form
\begin{equation}\label{}
    \int_{a}^{b}K(s,t)f(t){dt}=g(s),
\end{equation} a well known Fredholm equation of the first kind
\cite{Delves_Fredholm_1985} where $K(s,t)$ is the \textit{kernel}.
In general, such problems are \textit{ill-posed} where small
changes to the problem can make very large changes to the answer
obtained. The \textit{kernel} in \eqref{sup-input-distribution} is
analytic in $y$ over the whole plane for any $\nr$. However, the
right hand side of \eqref{sup-input-distribution} and its
derivative with respect to $y$ is infinite when
$y\,\rightarrow\,0$ for any $\nr$ and $\opvar_s$. Therefore,
\eqref{sup-input-distribution} does not provide a continuous
solution for ${p_{X}(x)}$ in which the $C_{\text{sup}}$ in
\eqref{sup-capacity-supremum-one} is attained. This leads us to
find a discrete input distribution in the form of
\begin{equation}\label{}
    {p_{X}(x)}=\sum_{i=1}^{N}{p_{i}}{\delta(x-x_{i})},
\end{equation} where $p_{i}$ and $x_{i}$ to be obtained solving
\begin{equation}\label{}
    g(s)=\sum_{i=1}^{N}{p_{i}}{K(s,t_{i})}.
\end{equation} The number of transmitters $\nt$ has no effect on
the result. The only requirement is to Euclidean norm of the input
vector $\bX$ be discrete irrespective of transmit diversity. For a
discrete input, we can pose a new optimisation problem to compute
the channel capacity
\begin{align}\label{cap-discrete}
    C_{\text{dis}}&={\text{sup}\atop{\Sigma_{i=1}^{N}p_{i}=1 \atop{\Sigma_{i=1}^{N}p_{i}{x_{i}^2}\,\leq \,\Pin}}}
                   \int_{0}^{\infty}\int_{0}^{\infty}p_{X}(x){p_{Y|X}(y|x)}\nonumber\\
                   &\times\,\log\left[{\frac{p_{Y|X}(y|x)}{\int_{0}^{\infty}p_{X}(x){p_{Y|X}(y|x)}}}\right]{dy}{dx}\nonumber\\
                   &={\text{sup}\atop{\Sigma_{i=1}^{N}p_{i}=1 \atop{\Sigma_{i=1}^{N}p_{i}{x_{i}^2}\,\leq \,\Pin}}}
                   \sum_{i=1}^{N}\int_{0}^{\infty}p_{i}{p_{Y|X}(y|x_i)}\nonumber\\
                   &\times\,\log\left[{\frac{p_{Y|X}(y|x_i)}{\sum_{j=1}^{N}p_{j}{p_{Y|X}(y|x_j)}}}\right]{dy},
\end{align} subject to the input power constraint $\Pin$. If the solution exists for \eqref{cap-discrete}, it will provide a good lower bound to
$C_{\text{sup}}$. However, this optimisation problem is extremely
difficult since the number of discrete points $N$ is unknown and
the optimum probabilities and their locations to be found
satisfying the input power constraint. Similar work is reported in
\cite{DTMRayleigh-Shami-2001} for a single antenna case and the
numerical evaluation is given using the Kuhn Tucker conditions in
order to verify the optimality of the capacity achieving mass
point probabilities and their locations. Further work is required
in this area to identify the optimal input at any SNR. The
capacity supremum in \eqref{sup-capacity-supremum-one}
can be treated as an upper bound for the capacity of non-coherent Rayleigh fading MIMO channel.\\

\subsection{Capacity for $\del\,>\,0$}\label{beta-positive-graph}
In addition to the capacity supremum found in the previous
subsection, we now elaborate another solution when $\del\,>\,0$ as
the capacity for a MIMO system with a small number of receive
antennas $\nr$. Since $\del$ is an increasing function of $\opvar$
for a given $\nr$ we can obtain the capacity
\begin{align}
C'_{\text{sup}}(\opvars)&=G(\opvars)-G(\nr),
\label{capacity-supremum-2}
\end{align} for the input power
\begin{equation}\label{input-power-2}
  \Pin=\frac{\opvars}{2{\nr}}{\exp\left[{{2}(\Psi(\nr)+\log{2})-\Psi(\opvars)}\right]}-1,
\end{equation} when $\del=\Psi(\nr)+\log{2}\,>0$.

\begin{figure}
\centering
  \includegraphics[width=\pictwidth]{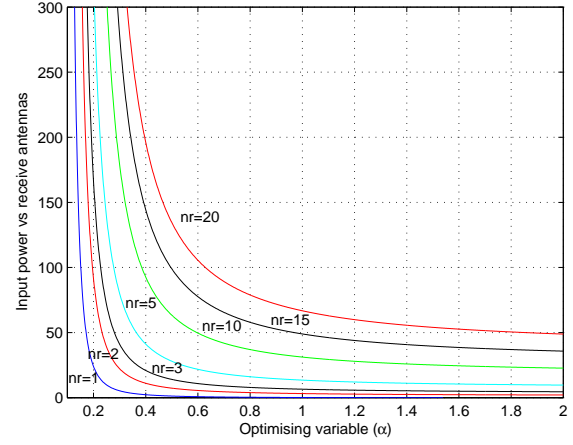}\\
  \caption{Input power \eqref{input-power-2} vs $\opvars$, showing that no solution exists when $\Pin=0$ for
           $\nr\,\geq\,2$.}\label{power-alpha}.
\end{figure}

The quantity $\opvars\,\in\,(0,\infty)$, and the upper limit of
$P$ is a function of $\nr$. For large $\nr$ and low input power
there is no $\opvars$ which provides a solution to
\eqref{input-power-2}. This could be seen from the asymptotic
value of \eqref{input-power-2}
\begin{equation}\label{input-power-2-limit}
    {\Pin{_{\text{asy}}}}=\frac{1}{{2{\nr}}}{\exp\left[{2(\Psi(\nr)+\log{2})}\right]}-1,
\end{equation} when $\opvars$ approaches infinity since the minimum $\Pin$ is
very high for $\nr\,\geq\,2$. Fig. \ref{power-alpha} illustrates
this, where there is no solution for $\Pin=0$ in
\eqref{input-power-2} when $\nr\,\geq\,2$.

This excludes us in finding a solution to
\eqref{capacity-supremum-2} for $\nr\,\geq\,2$. However, we can
show the capacity in \eqref{capacity-supremum-2} for $\nr=1$ when
the solution exists for $\opvars$ in the whole input range of
$\Pin$.

\begin{figure}
\centering
  \includegraphics[width=\pictwidth]{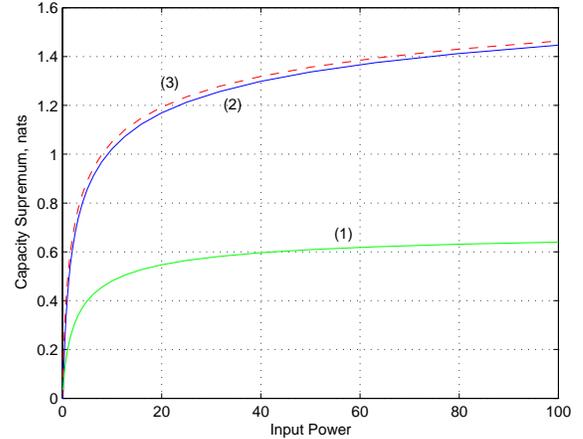}
  \caption{Comparison of capacities $C'_{\text{sup}}$ and
  $C_{\text{sup}}$ for $\nr=1$. (1) Capacity of a SISO
  non-coherent Rayleigh fading channel with a discrete input
  \cite{DTMRayleigh-Shami-2001}. (2) Capacity $C_{\text{sup}}$ when $\del=0$
  (3) Capacity $C'_{\text{sup}}$ when $\del=\Psi(\nr)+\log{2}$.}
\label{ch4-comparison}
\end{figure}
For $\nr=1$, \eqref{sup-capacity-supremum-one} and
\eqref{input-power-1} becomes
\begin{equation}\label{tarico-proof-cap}
    C_{\text{sup}}(\opvar_s)_{\nr=1}=(\opvars-\gamma-1)+\log{\Gamma(\opvar_s)}-\opvar_s{\Psi(\opvar_s)}
\end{equation} and
\begin{equation}\label{tarico-proof-power}
    \Pin{_{\nr=1}}=\opvar_s{e^{-\gamma-\Psi(\opvar_s)}}-1
\end{equation} giving the capacity bound shown in
\cite{Taricco-1997,Wenyi-Zhang-2005} for the non-coherent SISO
Rayleigh fading channel.

A direct upper bound to both \eqref{sup-capacity-supremum-one},
\eqref{capacity-supremum-2} can be drawn since both relate to
non-coherent capacity with channel coherent time $T=1$. The proven
capacity is inherently less than the capacity with perfect CSI.
Therefore we get
\begin{equation}\label{upper-bound}
    C'_{\text{sup}}(\opvars)\,<\,C_{\text{rcsi}}\,\,\,,\,\,C_{\text{sup}}(\opvar_s)\,<\,C_{\text{rcsi}}.
\end{equation}
In \cite{Siddharth-MIMO-2004}, the sublinear behavior of MIMO
channel capacity at low SNR is discussed. The two extremes, the
capacity of channels with full and partial CSI is studied
quantifying the maximum penalty for not having CSI at the
receiver. It is shown that capacity loss due to unknown CSI
increases monotonically with number of receive antennas,
elaborating the significance of CSI at the receiver. Therefore,
bounds shown in \eqref{upper-bound} become loose as $\nr$
increases.

\subsection{Asymptotic Analysis}
We consider the asymptotic analysis of the capacity supremum
\eqref{sup-capacity-supremum-one} when
$\Pin\,\rightarrow\,\infty$. $\Pin$ in \eqref{input-power-1}
approaches $\infty$ when $\opvar_s\,\rightarrow\,0$ since
\cite{Hand-book-mathematical-functions}
\begin{equation}\label{}
    {{\lim}\atop{{\opvar_s}_\to\,0}}\left({\frac{\opvar_s}{e^{\Psi(\opvar_s)}}}\right)\,\rightarrow\,\infty.
\end{equation}
Also when $\opvar_s=1$,
\begin{equation}\label{asy-input-power}
    \Pin=\frac{1}{{2}{\nr}}{\exp[\Psi(\nr)+\log{2}+\gamma]}-1,
\end{equation} where the input power is zero for $\nr=1$ and a non-zero quantity for $\nr\,>\,1$. Therefore, $\opvar_s=1$ is not valid
for $\nr\,>\,1$ and the $\opvar_s$ which produces $\Pin=0$ can be
found by solving
\begin{equation}\label{solver}
    \log{\opvar_s}-\Psi(\opvar_s)=\log({\nr})-\Psi(\nr),
\end{equation}
for each $\nr$. We find an expression for $\opvar_s$, when $\Pin$
approaches infinity, as a result of $\opvar_s$ approaching zero.
From \eqref{input-power-1} we get
\begin{align}
\log{\opvar_s}&=\Psi(\opvar_s)+\log{\nr}(1+\Pin)-\Psi(\nr).
\label{asymptotic-opvar}
\end{align}
Multiplying the both sides of \eqref{asymptotic-opvar} by
$\opvar_s$ and using
\begin{equation}\label{}
    {{\lim}\atop{{\opvar_s}_\to\,0}}\left({\opvar_s{\log{\opvar_s}}}\right)=0,\nonumber
\end{equation} and
\begin{equation}\label{}
    {{\lim}\atop{{\opvar_s}_\to\,0}}\left({\opvar_s{\Psi(\opvar_s)}}\right)=-1,\nonumber
\end{equation}
\cite{Hand-book-mathematical-functions}, we get
\begin{equation}\label{ch4-zetass}
    \opvar_s\,\approx\,\frac{1}{\log{\nr}(1+\Pin)}.
\end{equation} Substituting $\opvar_s$ from \eqref{ch4-zetass} into \eqref{sup-capacity-supremum-one} we get
the asymptotic capacity
\begin{equation}\label{Aysmtotic-capacity}
    C_{\text{sup}}\approx\log[\log{\nr}(1+\Pin)].
\end{equation}
Note here the double logarithmic behavior of the asymptotic value,
also depicted in \cite{Yingbin-nocsi-corre-2004,Taricco-1997,
Lapidoth-Moser-2002,Sengupta-2004}, for both the MIMO and SISO
configurations. The effect of capacity at high SNR is extensively
studied in \cite{Lapidoth-isit-2005} with the fading number
\begin{equation}\label{ch4-fading-number}
    \mathcal{X}\triangleq\,{{\lim}\atop{{\text{SNR}}\to\,\infty}}\left\{{C(\text{SNR})-\log\log{\text{SNR}}}\right\},
\end{equation} and degrees of freedom defined by
$n_{\text{min}}\triangleq\text{min}\{\nt,\nr\}$. The asymptotic
capacity
\begin{equation}\label{ch4-fn-asy-capacity}
    C(\text{SNR})=\log[1+\log(1+\text{SNR})]+\mathcal{X}+o(1),
\end{equation} indicates that at high SNR, capacity grows
double logarithmically where the antenna numbers hardly influence
the capacity. Our result further proves this argument since the
effect of $\nr$ in \eqref{ch4-fn-asy-capacity} on capacity is
negligible at high SNR.
\section{Numerical Results}\label{ch4-NR}
For $\nr=1$, the numerical results obtained with
$\del=\Psi(\nr)+\log{2}$  in Fig. \ref{ch4-comparison} is slightly
higher than $C_{\text{sup}}$ at any SNR. It is clear that $\del=0$
provides the lowest upper bound as predicted in Section
\ref{supremum-section}. However, for $\nr\,\geq\,2$, the bounds do
not exist for any SNR as explained in Section
\ref{beta-positive-graph}. The solutions to
\eqref{sup-capacity-supremum-one} and \eqref{input-power-1} lead
to good upper bounds for any $\nr$. Moreover, accurate bounds
could be realised using these solutions even at high SNR where
$\del=0$.

The SISO channel capacity shown by Abou-Faycal
\cite{DTMRayleigh-Shami-2001} with a discrete input is included in
Fig. \ref{ch4-comparison}. It is $\sim\,50\%$ lower than the
capacity supremum. Also capacity increase having more antennas at
the receiver is not promising at high SNR as in the coherent
Rayleigh fading MIMO channel.

\begin{figure}
\centering
  \includegraphics[width=\pictwidth]{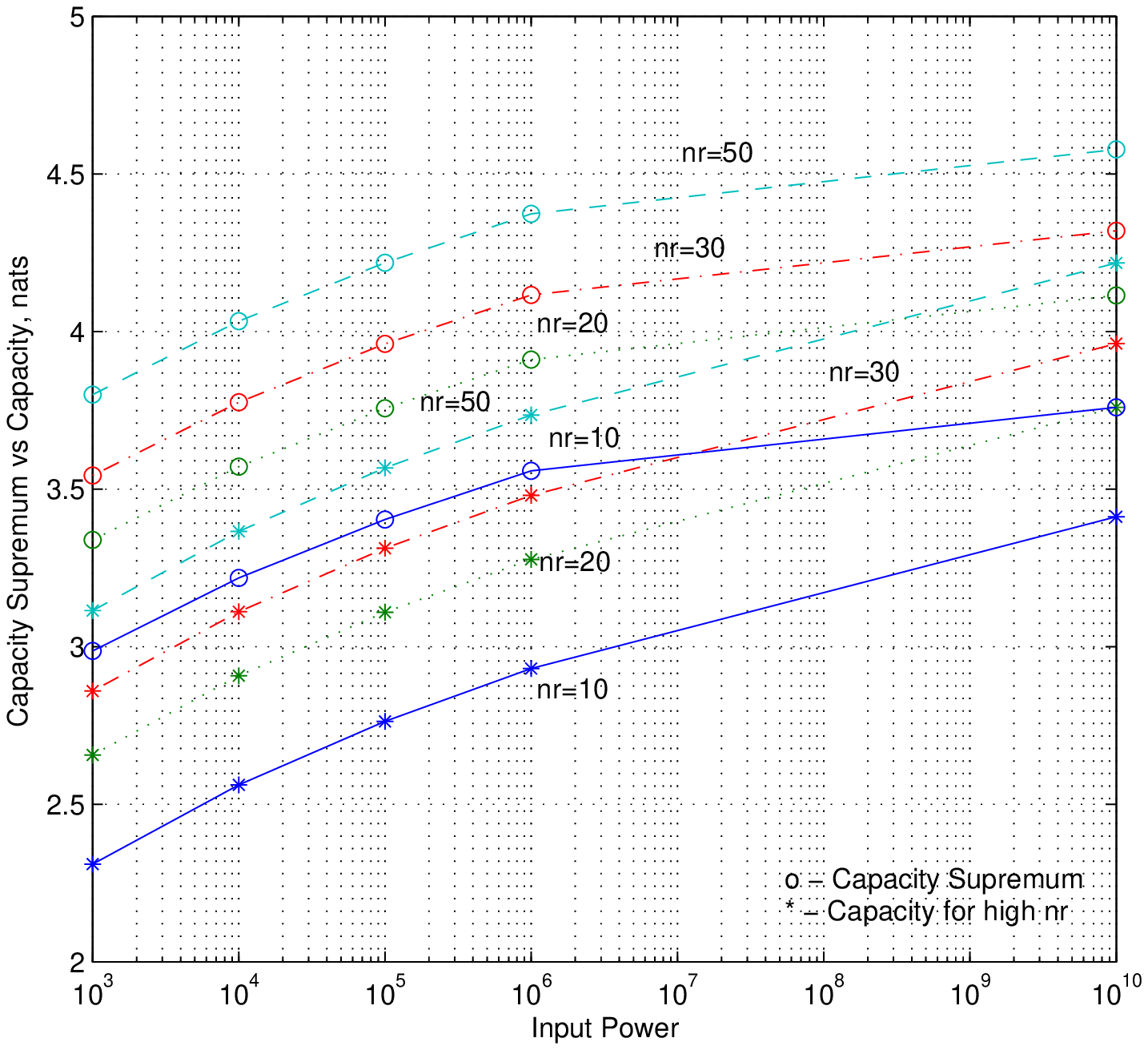}\\
  \caption{Capacity supremum in \eqref{sup-capacity-supremum-one} ``o'', vs the capacity for high $\nr$ given in
  \eqref{ch4-capacity-singupta} ``*'', at high input power. Both the capacities approach the limit shown in \eqref{Aysmtotic-capacity}
  resulting the double logarithmic asymptotic behavior.}\label{ch4-sing-mine}
\end{figure}

Furthermore, we compare the capacity supremum
\eqref{sup-capacity-supremum-one} with the results produced by
Sengupta and Mitra in \cite{Sengupta-2004} for high $\nr$ at high
$\Pin$ with $\nt=1$. Fig. \ref{ch4-sing-mine} depicts the
comparison of the capacity supremum vs capacity
\eqref{ch4-capacity-singupta} for $\nr\,\in\,\{10,20,30,50\}$ at
high SNR. From the graph, it is clear that the gap between the
capacities in \eqref{sup-capacity-supremum-one} and
\eqref{ch4-capacity-singupta} for a specific $\nr$ becomes small
as the input power increases. The limit is shown in
\eqref{Aysmtotic-capacity} for any $\nr$ illustrating the double
logarithmic behavior at high SNR.

\section{Conclusions}
\label{con} In this paper, we investigated the capacity of
uncorrelated MIMO channels when neither the receiver nor the
transmitter has the knowledge of CSI except its fading statistics.
Capacity supremum is given in two cases, where one predicts
accurate values for any number of receive antennas. The main
findings of this paper is the capacity supremum for the
non-coherent uncorrelated Rayleigh fading MIMO channel with no CSI
for a given number of receiver antennas at any SNR.

We have shown the asymptotic behavior of the capacity supremum at
high SNR. Furthermore, it is compared with the capacity shown in
the literature having large number of receive antennas.
Furthermore, the capacity supremum is shown to be independent of
number of transmit antennas and increases with the number of
receive antennas. The results of this paper can be used as an
upper bound to the non-coherent uncorrelated Rayleigh fading MIMO
channel with any input distribution.

The input distribution for the capacity upper bound is proven to
be non-continuous. Hence the optimisation problem is posed for a
discrete input with $N$ mass points. Capacity obtained from the
discrete input with optimal number of mass points, probabilities
and their locations will be a lower bound to the capacity supremum
derived in this paper. The capacity increase with increasing
number of receive antennas is not promising at high SNR as in the
coherent channel. Therefore, channel estimation becomes more
important as the number of antenna elements at the receiver is
increased.

\section{APPENDIX}
\subsection{ Derivation of the constraint in \eqref{constaints3}}
\label{constraint} We use \eqref{cond-pdf-mag1} in LHS of
\eqref{constaints3} to write
\begin{align}
    \int_{0}^{\infty}{p_{Y}(y)}{\log{y}}{d{{y}}}&=\int_{0}^{\infty}\int_{0}^{\infty}{p_{X}(x)}
    \left\{\frac{{y^{2{\nr}-1}}}{{2^{\nr-1}}{\Gamma(\nr)}}\right.\nonumber\\
    &\left.\times\,\frac{\exp\left[{-{\frac{y^2}{2{(1+x^2)}}}}\right]}{{(1+x^2)^{\nr}}}\right\}
    ({\log{y}})\,{d{{y}}}{d{{x}}}.
\label{App-constraint3}
\end{align}
Using the integral identity \cite{Table-of-integrals}:

\begin{align}
\int_{0}^{\infty}{x^t}{e^{-{u{x}^2}}}{\log{x}}{dx}&=\frac{u^{-\frac{t}{2}}}{4{\sqrt{u}}}
{\Gamma\left(\frac{t+1}{2}\right)}\nonumber\\
&\times\,\left[\Psi\left({\frac{t+1}{2}}\right)-\log{u}\right],
\end{align} we simplify \eqref{App-constraint3} as
\begin{align}
\int_{0}^{\infty}{p_{Y}(y)}{\log{y}}\,{d{{y}}}&=\frac{1}{2}\int_{0}^{\infty}{p_{X}(x)}\left[\Psi(\nr)\right.\nonumber\\
&\left.+\log{2}{(1+{x}^2)}\right]{d{{x}}}\nonumber\\
&=\frac{1}{2}[\del+\Psi(\nr)+\log{2}].
\label{proof-add-constraint}
\end{align} This constraint is used in addition to the constrained
input power in calculating the capacity supremum where
$\del=\E_{x}\{\log(1+x^2)\}$.

\section{ACKNOWLEDGEMENTS}
National ICT Australia (NICTA) is funded through the Australian
Government's \emph{Backing Australia's Ability Initiative}, in
part through the Australian Research Council. The authors would
like to thank Michael Williams for helpful discussions.


\end{document}